\documentclass[submission, Proceedings]{SciPost}

\hypersetup{
    colorlinks,
    linkcolor={red!50!black},
    citecolor={blue!50!black},
    urlcolor={blue!80!black}
}

\usepackage[bitstream-charter]{mathdesign}
\DeclareSymbolFont{usualmathcal}{OMS}{cmsy}{m}{n}
\DeclareSymbolFontAlphabet{\mathcal}{usualmathcal}

\begin{document}

\begin{center}{\Large \textbf{
Probing gluon density fluctuations at large momentum transfer $|t|$ at HERA\\
}}\end{center}

\begin{center}
Arjun Kumar$^\textsubscript{*}$
\end{center}

\begin{center}
{\bf } Indian Institute of Technology Delhi, NewDelhi-110016, India
\\

*arjun.kumar@physics.iitd.ac.in
\end{center}

\begin{center}
\today
\end{center}


\definecolor{palegray}{gray}{0.95}
\begin{center}
\colorbox{palegray}{
  \begin{tabular}{rr}
  \begin{minipage}{0.1\textwidth}
    \includegraphics[width=22mm]{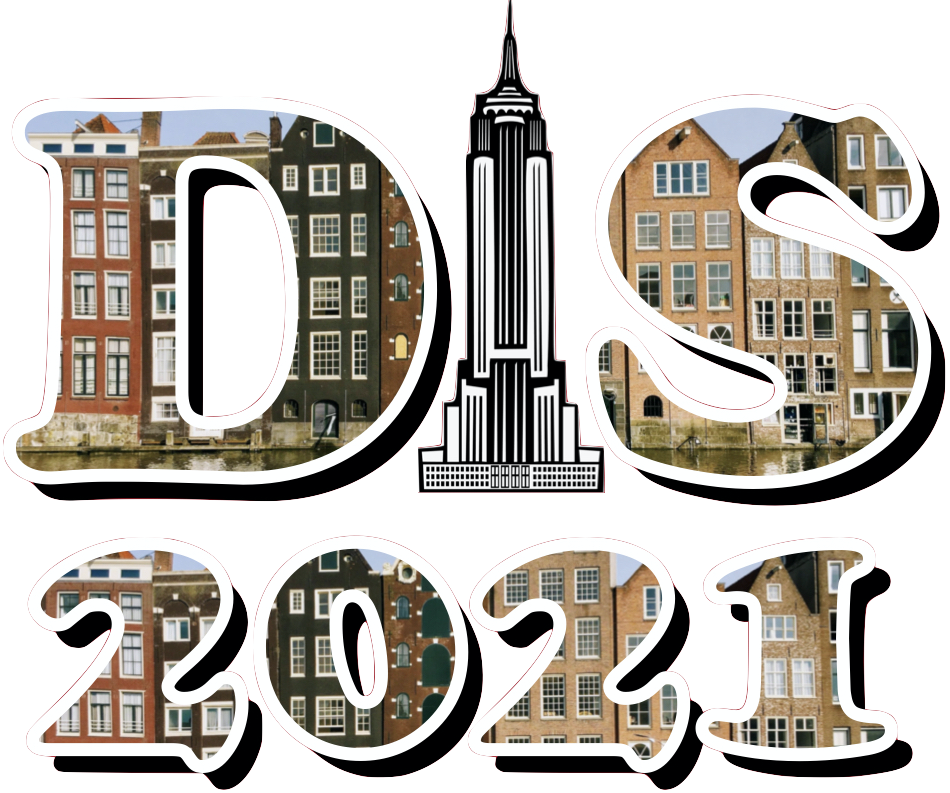}
  \end{minipage}
  &
  \begin{minipage}{0.75\textwidth}
    \begin{center}
    {\it Proceedings for the XXVIII International Workshop\\ on Deep-Inelastic Scattering and
Related Subjects,}\\
    {\it Stony Brook University, New York, USA, 12-16 April 2021} \\
    \doi{10.21468/SciPostPhysProc.?}\\
    \end{center}
  \end{minipage}
\end{tabular}
}
\end{center}

\section*{Abstract}
{\bf
The information on the gluonic structure and its fluctuations is captured by the differential $|t|$ spectrum in diffractive events. The incoherent cross-section is sensitive to the fluctuations in the target wavefunction in such events. We investigate the incoherent $ep$ cross section in $J/\psi$ photoproduction using the impact-parameter dependent dipole model. The spatial gluonic structure is modelled as hotspots of gluon density having substructure where this substructure is modelled as hotspots within hotspots. We find that three levels of the substructure provide a good description of all the data, available up to $|t|=30~$GeV$^2$. We investigate these fluctuations in both the saturated and non-saturated dipole models and compare our predictions with the HERA Data. 
}

\vspace{10pt}
\vspace{10pt}

\section{Introduction}
\label{sec:intro}
Diffractive events such as Exclusive Vector Meson production \cite{Aktas:2005xu,Chekanov:2002xi,Alexa:2013xxa} in $ep$ scattering serve as an excellent probe to investigate the transverse structure of proton. At low momentum transfer $|t|$, the differential cross-section is dominated by the coherent cross-section where the proton stays intact while at large $|t|$ the incoherent cross-section dominates where the proton breaks up. In the Good-Walker formalism \cite{Good:1960ba}, the coherent cross-section probes the average profile of the target while the incoherent cross-section is sensitive to the fluctuations in the target wavefunction. In $eA$ collisions, event-by-event fluctuations in nucleon positions \cite{Toll:2012mb} has already been utilised to study the incoherent cross-section at an Electron-Ion Collider \cite{Accardi:2012qut,AbdulKhalek:2021gbh}. At small-x or high energy one works in the dipole or the target rest frame  \cite{ GolecBiernat:1999qd, Kowalski:2003hm, Kowalski:2006hc } where the incoming virtual photon first splits into a quark-antiquark pair forming a dipole which subsequently interacts with the target via strong interaction by exchanging vacuum quantum numbers. The transverse coordinates of the dipole are frozen during  the interaction thus it is advantageous to work in impact parameter space and one can measure the momentum exchange at the proton vertex to get the information about spatial structure. In 2016, M\"antysaari and Schenke included the fluctuations in the saturated dipole model by assuming the profile of the proton to be made of up of three hotspots of gluon density to explain the incoherent cross-section but the shortcoming  of the hotspot model is that it is valid in the low momentum transfer region, where $|t|$ < 2.5 GeV$^2$  \cite{ Mantysaari:2016ykx,Mantysaari:2016jaz}. The model also underestimates the coherent cross section, as well as the very small |t| region of the incoherent spectrum. Here we propose a refined hotspot model \cite{Kumar:2021zbn} which explains the incoherent data at both low and high momentum transfer. We investigate these fluctuations in both the saturated and non-saturated dipole models and compare our predictions with the HERA Data.

\begin{figure*}[t]
	\centering
	\includegraphics[width=0.315\linewidth]{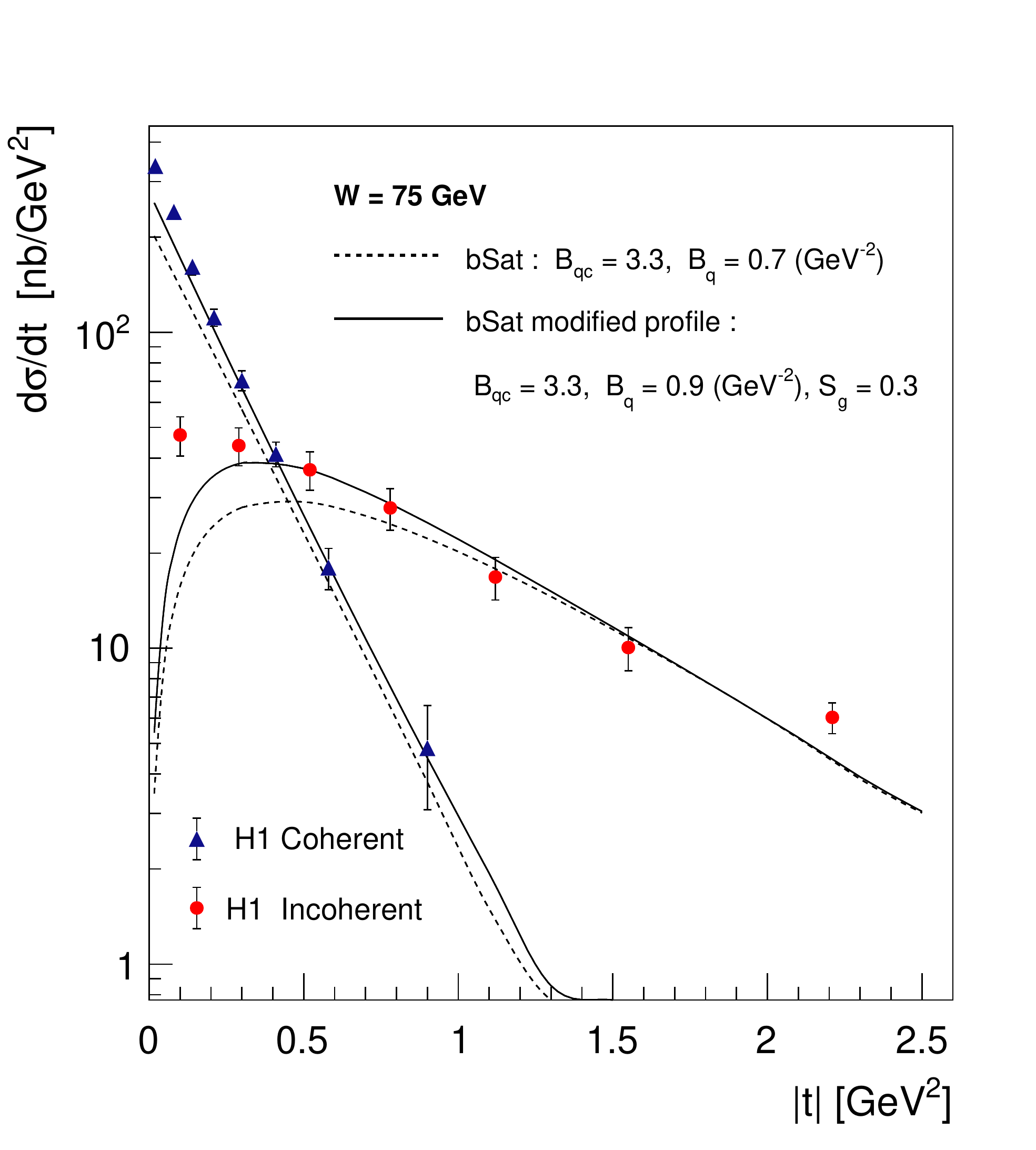}\hskip0.5cm
	\includegraphics[width=0.58\linewidth]{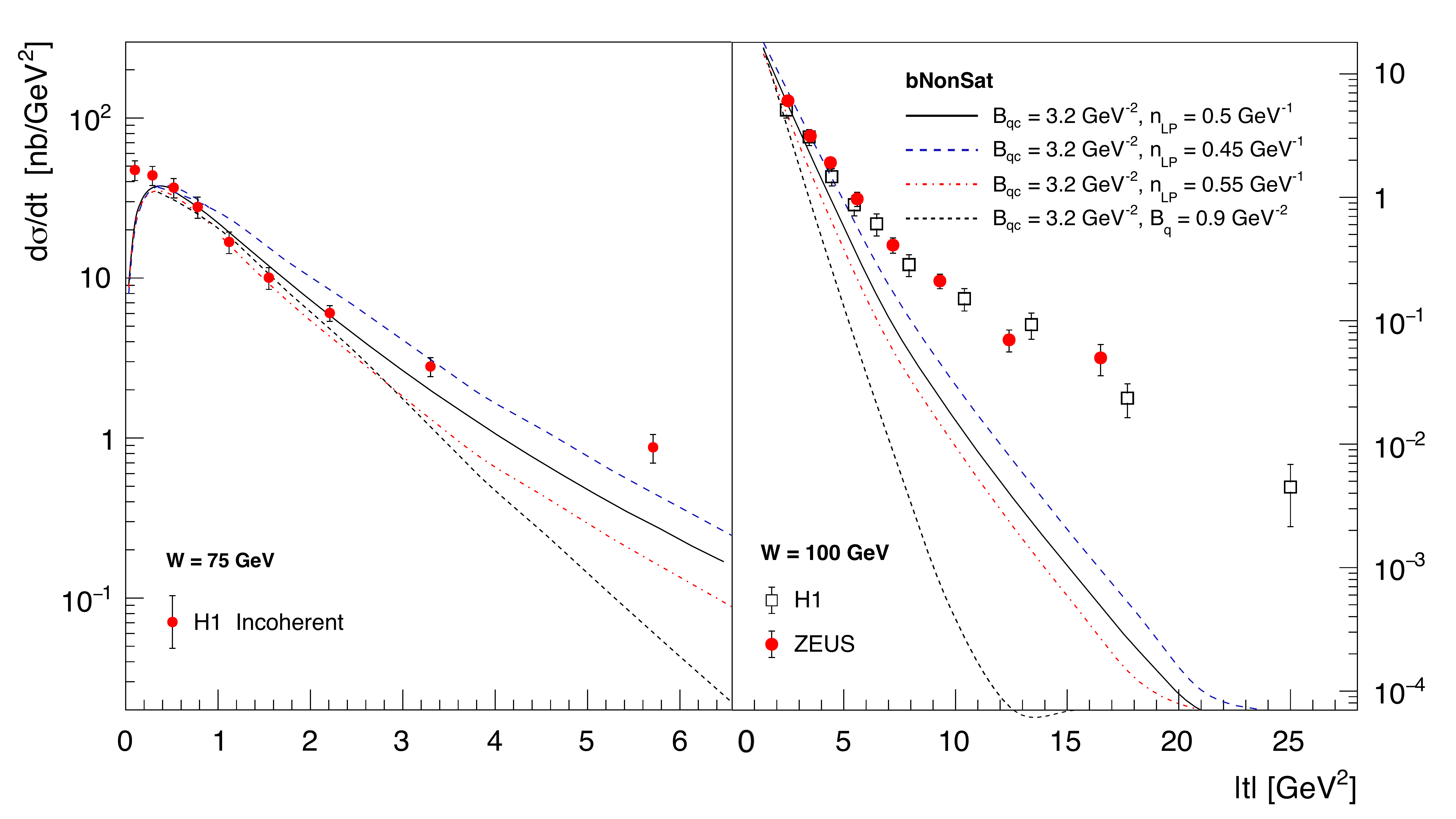}
		\caption{The $|t|$ dependence of $J/\Psi$ photoproduction \cite{Alexa:2013xxa,Aktas:2003zi,Chekanov:2009ab} in the hotspot model with the gaussian profile \& modified profile in bSat model on left panel and with gaussian \&  laplace profile in bNonSat model on the right panel. }
	\label{laplace}
\end{figure*}
\section{The Color Dipole Model }
At high energy the amplitude factorizes and is given by the convolution of three sub-process, first the photon splits into quark-antiquark pair forming a dipole, then the dipole interacts with proton elastically via strong interaction and finally the dipole forms the final state. The scattering amplitude is given by,
\begin{align}
\mathcal{A}_{T,L} (x_{\mathbb{P}},Q^2,\textbf{$\Delta$})=i\int d^2 \textbf{r}\int d^2\textbf{b}\int \frac{dz}{4 \pi}   (\Psi^*\Psi_V)_{T,L}(Q^2,\textbf{r},z)  e^{-i[\textbf{b}-(1-z)\textbf{r}].\textbf{$\Delta$}} \frac{d\sigma _{q\bar{q}}}{d^2\textbf{b}}(\textbf{b},\textbf{r},x_\mathbb{P})
 \end{align}
 We consider two versions of the dipole cross-section, the saturated dipole model (bSat) and the non-saturated dipole model (bNonSat). The bSat model dipole cross-section is given by \cite{Bartels:2002cj}:
 \begin{eqnarray}
 \frac{d\sigma _{q\bar{q}}}{d^2\textbf{b}}(\textbf{b},\textbf{r},x_\mathbb{P})=
 2\big[1-\text{exp}\big(-\frac{\pi^2}{2N_C} \textbf{r}^2 \alpha_s(\mu^2) x_\mathbb{P} g(x_\mathbb{P},\mu^2) T_p(\textbf{b})\big)\big]
 \end{eqnarray}
 The linearised version is called the bNonSat model \cite{Mantysaari:2018nng}:
 \begin{equation}
 \frac{d\sigma _{q\bar{q}}}{d^2\textbf{b}}(\textbf{b},\textbf{r},x_\mathbb{P})=\frac{\pi^2}{N_C}\textbf{r}^2\alpha_s(\mu^2) x_\mathbb{P} g(x_\mathbb{P},\mu^2)  T_p(\textbf{b})
 \end{equation}
 the transverse profile of the proton in both the models is usually assumed to be Gaussian where $T_p(\textbf{b}) = \frac{1}{2 \pi B_G}\exp\bigg(-\frac{\textbf{b}^2}{2B_G}\bigg)$. 
 When we consider the fluctuations (due to different number of interacting constituents or transverse positions of constituents) in target wavefunction the incoherent cross-section in the Good-Walker formalism is given by the difference in the second moment and first moment of the amplitude and given by, 
  \begin{equation}
  	\frac{d \sigma_{\rm incoherent}}{dt} = \frac{1}{16 \pi}\bigg(\big< \big| \mathcal{A}(x_{\mathbb{P}},Q^2,\textbf{$\Delta$})\big|^2\big>_\Omega ~ ~- ~~\big| \big<\mathcal{A}(x_{\mathbb{P}},Q^2,\textbf{$\Delta$})\big>_\Omega\big|^2\bigg)
  \end{equation}
    
  \begin{figure}[t]
  	\centering
  	\includegraphics[width=0.5\linewidth]{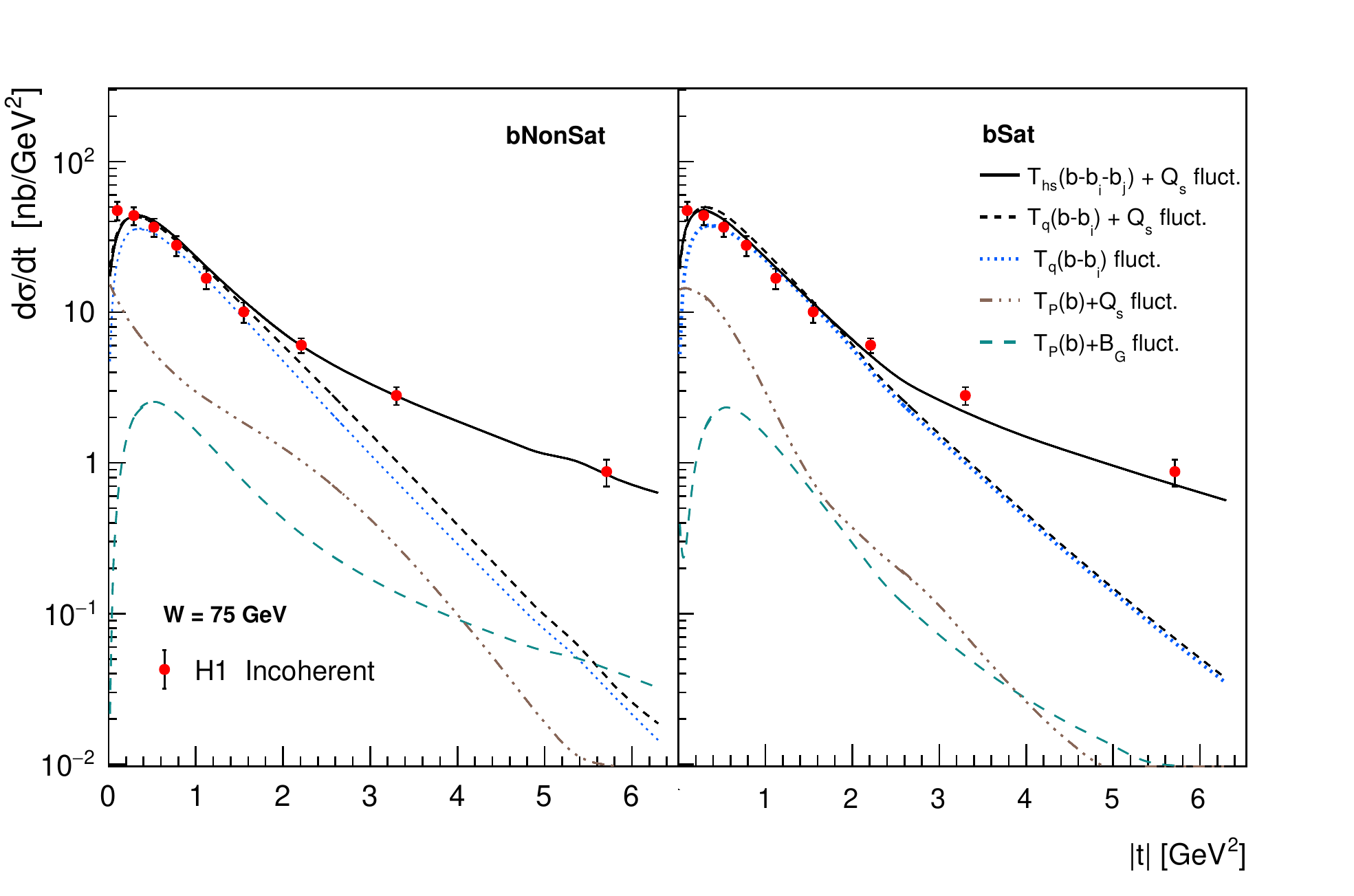}\hskip-0.3cm
  	\includegraphics[width=0.5\linewidth]{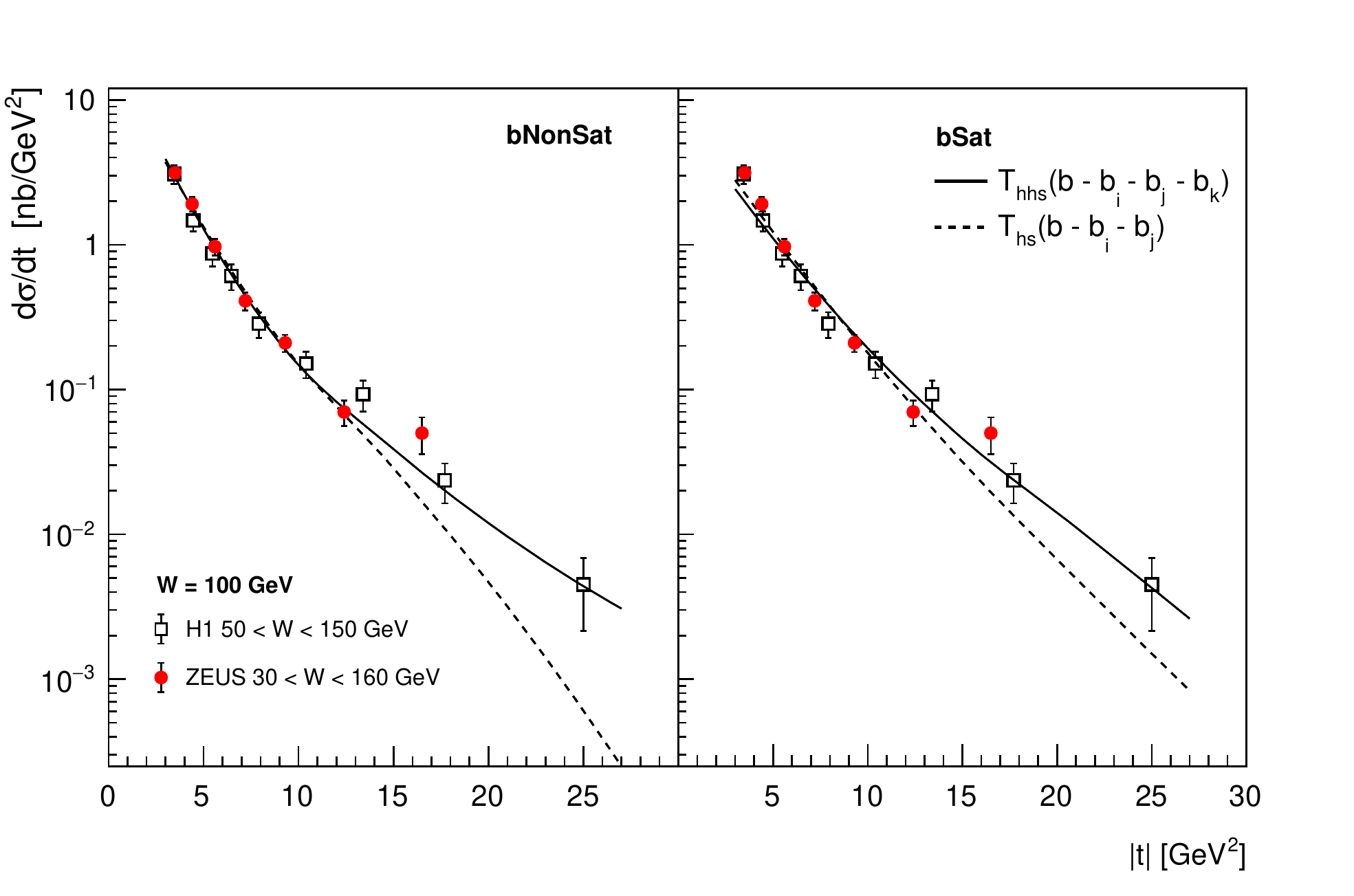}
  	\caption{The $|t|$ dependence of incoherent $J/\Psi$ photoproduction in refined hotspot model.}
  	\label{master}
  \end{figure}
\begin{figure}[t]
	\centering
	\includegraphics[width=0.4\linewidth]{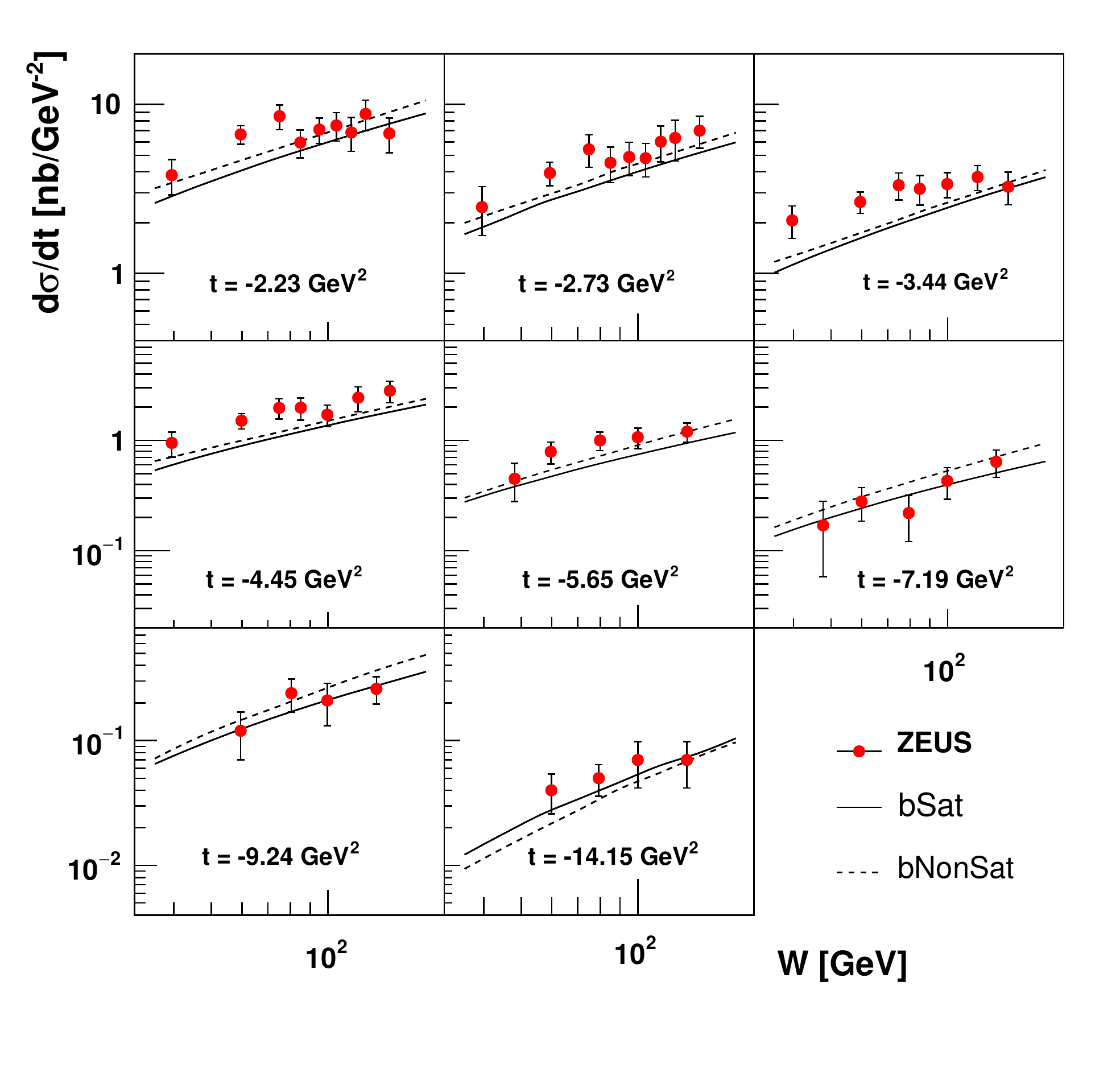} \hspace{1.5cm}
		\includegraphics[width=0.36\linewidth]{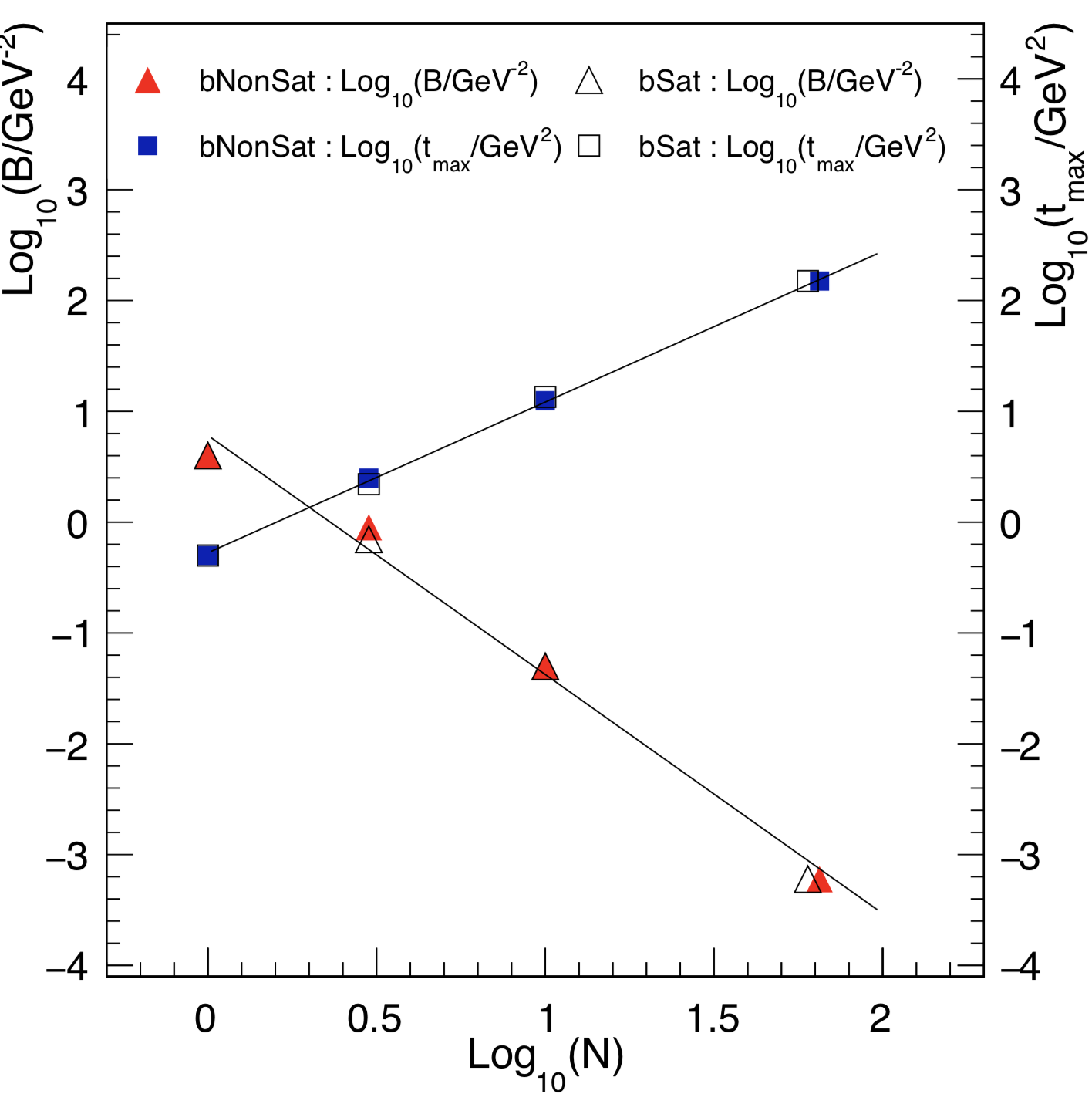} 
	\caption{Left : Energy dependence of the incoherent J/$\Psi$ production in the saturated and non-saturated version of the refined hotspot model, Right : Scaling behaviour of the spatial gluon density fluctuations inside the proton of the parameters B and N. }
	\label{energy}
\end{figure}

  Now on including fluctuations \cite{Mantysaari:2016jaz,Traini:2018hxd} the profile in the hotspot model becomes $	T_p(\textbf{b}) \rightarrow\\ \sum_{i=1}^{N_q}T_q(\textbf{b-b$_i$}) $ and the shape of the hotspots is  $	T_{q}(\textbf{b}) = \frac{1}{2 \pi B_{q}}\exp\big[-\frac{\textbf{b}^2}{2B_{q}}\big]$ where  $B_q$ is the width of hotspots and $\textbf{b}_i$ are sampled from a Gaussian width $B_{qc}$, In the bSat case, with the Gaussian hotspots, the coherent data and the incoherent data at very small |t| is underestimated (see Fig.~\ref{laplace}  left panel) so we choose a modified profile which retains the coherent cross-section. The modified profile function is given by $	T_q(\textbf{b}) = \frac{1}{2 \pi B_q}\frac{1}{\big(\exp\big[\frac{\textbf{b}^2}{2B_q}\big]-S_g\big)}$, the  parameter $S_g$ can be interpreted as being related to probed correlations between gluons within the hotspots. With this modified profile we get a good description of the coherent and incoherent data at low $|t|$ as depicted in the left panel of Fig.~\ref{laplace}. The hotspot model also fails at large momentum transfer as illustrated on the right  panel in Fig.~\ref{laplace} where we plot the incoherent cross-section using bNonSat model with Gaussian ( $	T_{q}(\textbf{b}) = \frac{1}{2 \pi B_{q}}\exp\big[-\frac{\textbf{b}^2}{2B_{q}}\big]$) and Laplace ($T_{q}(\textbf{b}) = \frac{1}{4 \pi n_{LP}^3} b K_1\big[-\frac{b}{n_{LP}}\big]$) shape of the hotspots and we see both the profiles largely underestimates the data at large momentum transfer $|t|$. An important observation in the data is the change of slope which signifies that the dipole interacts with different size objects inside the proton and this is the motivation of the refined hotspot model (refer \cite{Kumar:2021zbn})where we assume that the three hotspots consists of substructure i.e the bigger hotspots are made up of smaller hotspots of gluon density. In the refined hotspot model with geometrical event-by-event fluctuations in proton wavefunction, the profile function is given by \cite{Kumar:2021zbn} : 
  \begin{eqnarray}
  T_{P}(\textbf{b}) &=&
  \frac{1}{2 \pi N_q N_{hs}N_{hhs} B_{hhs}} 
  \sum_{i}^{N_q}\sum_{j}^{N_{hs}}\sum_{k}^{N_{hhs}}
  e^{-\frac{(\textbf{b}-\textbf{b}_i-\textbf{b}_j-\textbf{b}_k)^2}{2B_{hhs}}}
  \end{eqnarray}
  Here, \textbf{b}$_i$, \textbf{b}$_j$ and \textbf{b}$_k$ determine the transverse positions of the larger, smaller and smallest hotspots respectively, which fluctuates event-by-event.  $N_q$, $N_{hs}$ and $N_{hhs}$ are the number of  hotspots in each level of substructure respectively and $B_{hhs}$ is the width of the smallest hotspots.  A detailed information on the refined hotspot model  can be found in \cite{Kumar:2021zbn}. As illustrated in Fig.~\ref{master}, we obtained a good description of the data with two layer of substructure upto $|t|$ $\leq$ 13$~ GeV^{2}$ and to explain the whole $|t|$ spectrum we need three layers of substructure with large event-by-event fluctuations. In the refined hotspot model we obtained the convergence  in around 1800 configurations. Also for the bSat case we use the modified profile in order to retain the coherent cross-section. For the large momentum transfer region, it is the size of smallest fluctuating object which is essential. We also obtained a good description of the energy dependence of the incoherent data as illustrated in left panel of Fig.~\ref{energy}. 

\noindent In our model, the size and number of hotspots are highly correlated such that the logarithm of the number of hotspots and their size fall on a line as shown in right panel of Fig.~\ref{energy}. We also indicate the upper value of $|t|$ for which the fluctuations at a scale B contributes, these are the t- values at which the incoherent spectrum changes slope.
\section{Conclusion }
We have shown that the refined impact parameter dependent hotspot dipole model is an excellent tool to study the gluonic structure and its fluctuations of the proton using J/$\psi$ photoproduction at HERA. We also observed that the substructure  of hotspots manifest itself in the large $|t|$ part in $|t|$ dependence and it is the size of smallest fluctuating object which is important at large momentum transfer. In order to explain the whole $|t|$ spectrum we require three layers of substructure. Also we have shown that a modified thickness function with an extra parameter in bSat case leads to good description of coherent data and incoherent data at very low $|t|$, this extra parameter can be interpreted as a measure of correlations among the gluons within the hotspots.  We also obtained a good agreement with the energy dependence with both the saturated and non-saturated versions of the refined hotspot model. Additionally  their is a scaling in the structure of transverse gluon density fluctuations in the proton in our model such that the logarithm of the number of hotspots and their size fall on a line. If this scaling persists, we expect our model with three levels of gluonic substructure in the proton will be relevant for $|t|\leq150~GeV^2$.  \\

\section*{Acknowledgements} 
 This work was done in collaboration with Tobias Toll. We thank T. Ulrich, T. Lappi, B. Schenke and H. M\"antysaari for useful discussions. The work of A. Kumar is supported by Department of Science \& Technology, India under DST/INSPIRES/03/201 8/000344. We have used computing resources of our HEP-PH group at IIT Delhi.




\bibliography{bibliography}
\nolinenumbers

\end{document}